\documentstyle[preprint,prl,aps,psfig]{revtex}
\newcommand{\beq}{\begin{equation}}
\newcommand{\eeq}{\end{equation}}
\newcommand{\beqn}{\begin{eqnarray}}
\newcommand{\eeqn}{\end{eqnarray}}
\newcommand{\vf}{\varphi}
\newcommand{\s}{\sigma}

\newcommand{\fr}{\frac}

\begin{document}

\begin{titlepage}
\baselineskip 24pt plus 2pt minus 2pt
\rightline{ROME1 prep. 96/1161}
\vspace*{0.2cm}

\begin{center}{ \Large \bf Renormalons in the effective potential of the  
vectorial $(\vec{\phi}^{2})^{2}$  model}
\end{center}

\centerline{\bf Luca Galli, Ignazio Scimemi}
\centerline{Dip. di Fisica, Univ. ``La Sapienza", P.le A. Moro,
I-00185 Rome, Italy}
\centerline{e-mail:\hspace{.5cm}
 galli@vxrm64.roma1.infn.it, scimemi@vxrm64.roma1.infn.it}
\begin{abstract}
We study the properties of ultraviolet renormalons in the vectorial
$(\vec{\phi}^{2})^{2}$  model.
This is achieved by studying 
 the effective potential of the theory at next to leading 
 order of the $1/N$ expansion,   the appearence of
the renormalons in the perturbative series
and  their  relation  to the imaginary part of the potential.
We also consider the mechanism of renormalon cancellation
 by ``irrelevant"  higher  dimensional  operators.
\end{abstract}

\end{titlepage}

\section{Introduction}
The $(\vec{\vf}^2)^2$ theory  and QED  suffer from
non-perturbative singularities at large  energy scales \cite{landau}.
This  is a common behaviour of renormalizable  ``infrared free" quantum field 
theories in four dimensions and it is a consequence of the presence
 of the so-called
Landau ghost in the running coupling constant.
 These theories are considered ``trivial": 
renormalization group arguments  tell us that  
 there is no consistent
way perturbatively to take the large cut-off limit of the regularized theory
without driving to zero the renormalized coupling constant \cite{wilsonkogut74}.
Another problem is that  the  next to leading $1/N$ order effective
potential of the vectorial $(\vec{\phi}^{2})^{2}$
 model  is complex \cite{root}  for the  same reason as that for the
Landau ghost, i.e.  the ultraviolet behaviour of the running coupling
constant.
All these aspects are  connected to the behaviour
of  high order terms  in the perturbative
 expansion. The perturbative  series may indeed  diverge due to
two different reasons:
 
1)   proliferation in the number of Feynman graphs.
 The number of diagrams  that contribute to 
a fixed perturbative order $n$  grows
as $n!$ (instanton contributions);

2)  pathological behaviour associated to single diagrams.
   The renormalized amplitude
of single graphs that at the $n$-th  perturbative
order have a number of subdivergencies of  order $n$ 
goes as    $ n!$ (renormalon contributions)  \cite{lautrup}.

 The $1/N$ expansion  allows us  to resum  the second 
class of diagrams  which are those  studied in this paper.
It  has been argued  by \cite{'tHooft} that in this case their contribution 
is the dominant one.
 The Borel trasform method   provides  us a general 
recipe to treat  divergent series \cite{dingle},
when the Borel trasform exists.
According to this method, given some 
physical quantity whose  
perturbative series in the coupling constant ``$g$" is
\beq
F[g]=\sum_{n=1}^{\infty} F_{n}\, g^{n}=F_{1}\, g+F_{2}\, g^{2} + \, ... \, ,
\eeq
 one defines its  Borel trasform  by 
\beq
B[F[g],t]= \sum_{n=0}^{\infty} F_{n+1}\, \fr{t^{n}}{n !}=
F_{1}\, +F_{2}\, t + F_{3}\, \fr{t^{2}}{2} +\, ... \; . 
\label{borel}
\eeq
Then, formally
\beq
F[g]=\int_{0}^{\infty} \, dt e^{-t/g}\, B[F[g],t]\, ,
\label{inter}
\eeq  
in the sense that the expansion for $B[F[g],t]$ reproduces the expansion for 
$F[g]$ term by term.
The inverse trasform in eq.~(\ref{inter}) is 
 meaningfull if the integral of  converges
and  $B[F[g],t]$  has no singularities in the range of integration.
 Unfortunatly  the $(\vec{\vf}^2)^2$ theory, QED and QCD 
are not even Borel  summable \cite{parisi,all}.
In these theories the Borel trasforms of physical quantities have
 singularities on the positive axis of Borel  plane.
In order to do the integral (\ref{inter}), one should then
 specify a prescription
to go around the singularities. Unfortunately
 the result  will depend  on the 
prescription which has been chosen.
The  singularity closest to the origin in $t$,
which we denote as ``first renormalon",  gives the leading contribution
to the ambiguity of the perturbative series. 

 Parisi \cite{parisi} conjectured that UV renormalon ambiguities
in the $n$-point 1-PI Green functions  are proportional to the 
insertion  
of local ``irrelevant" composite operators in these  Green functions. 
The first renormalon can be removed
 considering the full set of the  operators of dimension six
 ($\vf^{6} ,\vf^{2}\partial_{\mu} \vf\partial^{\mu} \vf,...$),
the second renormalon considering operators of dimension eight and so on.

In this paper we compute the first renormalon contribution to  the effective 
potential at the next-to-leading order (NLO) in  $1/N$ and  check
whether the conjecture of Parisi applies.
Since at the order $1/N$ at which we work,
 the  effective potential is built using
 all zero momentum 1-PI Green funtions
 the renormalon contributions are completly removed with the help of higher 
dimensional operators  without derivatives.
We  show indeed tha t in order to remove the first renormalon  in the 
effective potential it is sufficient to add to the Lagrangian the operator 
$(\vec{\phi}^{2})^{3}$ in agreement with the conjecture of Parisi.
Thus, the fact  that the NLO effective potential develops an imaginary part 
does not
mean at all that the theory is inconsistent.
The Borel trasform and the renormalon theory  give us a method to  isolate
this imaginary contribution and cancelling it with the insertion of
higher dimensional operators.

This paper is organized as follows: in section II
 we recall the basic ingredients for the calculation
 of the effective  potential and introduce the notation;
 in section III we compute the Borel trasform of the potential;
  in section IV we discuss its  renormalization.
 In section V  then, we  isolate the first renormalon contribution
to the effective potential and
we compare the result to  Parisi's  hypothesis
on  1-PI  green functions.
The conclusion can be found in section VI.

\section{The effective potential  in the leading and next-to-leading  $1/N$
expansion }

In this section we give the basic elements and fix the notation of the
$1/N$ epansion of the effective potential.
Our derivation   follows  closely those which can be found
in the past literature on this subject \cite{root,coleman}.
The starting point is a $(\vec{\phi}^{2})^{2}$-vector  theory,
with N scalar fields $\phi_{a}$. 

The lagrangian density, after the introduction 
 of a constraint-field (~lagrangian multiplier~) $\hat{\chi}$, 
can be written as

\beqn
\nonumber
\cal{L} &=&\fr{1}{2} \partial_{\mu} \phi_{a} \partial_{\mu} \phi_{a}+
\fr{m_{0}^{2}}{2} \phi_{a} \phi_{a}+
\fr{g_{0}}{8} (\phi_{a} \phi_{a})^{2} 
-\fr{1}{2g_{0}} \left( \fr{\hat{\chi}}{\sqrt{N}} -
\fr{g_{0}}{2} \phi_{a} \phi_{a}-
m_{0}^{2} \right)^{2}=
\\ 
&=&\fr{1}{2} \partial_{\mu} \phi_{a} \partial_{\mu} \phi_{a} -
\fr{\hat{\chi}^{2}}{2 g_{0} N} +\fr{\hat{\chi}}{\sqrt{N}}\phi_{a} \phi_{a}+
\fr{m_{0}^{2}}{g_{0}} \fr{\hat{\chi}}{\sqrt{N}} \, .
\eeqn

 The fields $\phi_{a}$ and the constraint
 $\hat{\chi}$ can be  expressed as their  vacuum expectation value 
($\phi$ and ${\chi}$ respectively)
plus quantum fluctuations.
By taking only the first component of the fields $\phi_{a}$
to have a non-zero vacuum expectation value ( VEV ), we have

\beqn
\begin{array}{ll}
\phi_{1} (x)=\s (x) + \vf \; \sqrt{N} \, , & \hspace{1cm}
\hat{\chi} (x)= \tilde{\chi} 
(x) +\chi \; \sqrt{N} \, ,
\\
\phi_{a} (x) =\vf_{a} (x) \, , & \hspace{1cm}  a=2,3,...,N\;   .
\end{array}
\label{miao}
\eeqn

Up to an irrelevant costant and linear terms,
the lagrangian  is then given  by

\beq
\cal{L}=\fr{1}{2} \partial_{\mu} \vf_{a} \partial_{\mu} \vf_{a}+
\fr{\chi}{2}\vf_{a} \vf_{a}+
\fr{1}{2} \partial_{\mu} \s \partial_{\mu} \s+
\fr{\chi}{2}\s^{2} +\vf \s \tilde{\chi} -\fr{\tilde{\chi}^{2}}{2 N g_{0}}+
\fr{\tilde{\chi}}{2 \sqrt{N}} (\s^{2} +\vf_{a} \vf_{a})\, ,
\label{lag}
\eeq
where the sum over repeated  Latin indices goes now from 2 to N. 

\subsection{Basic equations and their tree level solution}

The effective potential $V(\vf ,\chi)$ of the modified theory eq.~(\ref{lag})
reduces  to the  effective potential of the original theory
 when $\chi$ satisfies the requirement 

\beq
\left. \fr{\partial V(\vf ,\chi)}{\partial \chi}\right|_{\vf}=0 \, .
\label{www}
\eeq
\noindent
 We use (\ref{www}) to define $\chi $ as a function of $ \vf $. 

In order to renormalize  the potential, we impose the following 
renormalization conditions

\beqn
\nonumber
& & \left. \chi(\vf) \right|_{\vf=0}=\mu^{2} \, ,
\\ \nonumber 
& & \left.
\fr{d \, V(\vf ,\chi(\vf))}{d \,\vf^{2}}\right|_{\vf=0,\;\chi=\mu^{2}}=
\fr{\mu^{2}}{2} \, ,  
\\ 
& & \left.
\fr{d^{2} \, V(\vf ,\chi(\vf))}{d^{2} \,\vf^{2}}\right|_{\vf=0,\;\chi =
\mu^{2}}=
\fr{g \, N}{4} \, , \label{renc}
\eeqn
where $\mu^{2}$ and $g$ are the renormalized mass and coupling 
constant.

Finally, we introduce the expantion of the potential in powers of $N$
\beq
V=N \; V_{L}+ V_{NL}+... \, ,
\label{mare}
\eeq
where $ N \, V_{L}$ is of order $N$, $ V_{NL}$ of order 1,  etc.

A similar expansion holds also for $\vf$ and $\chi$. 
Thus, for example we  may write the solution
of eq.~(\ref{www}) as
\beq
\chi=\chi_{0}+\fr{1}{N}\chi_{1}+...
\eeq
 At lowest order in $1/N$, $\chi_0$ is the solution of
\beq
\left.
\fr{\partial V_{L}}{\partial \chi}\right|_{\chi =\chi_{0}}=0\, . \label{czero} 
\eeq

\begin{figure}
\vskip2cm
\centerline{\hbox{\psfig{figure=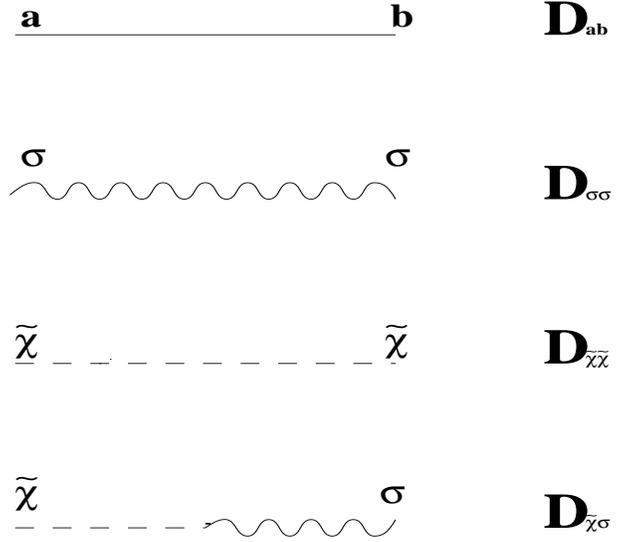,height=7cm,width=8cm,angle=0}}}
\vskip1cm
\caption{Propagators for the theory involving the $\tilde{\chi}$ field.}
\label{propa}
\end{figure}

  It is straightforward to derive the tree level vacuum potential, expressed
in terms of the fields $\vf$ and $\chi$

\beqn
V_{tree}=-\fr{\chi^{2}}{2 g_{0}}+ \fr{N}{2} \chi \vf^{2}+
\fr{m_{0}^{2}}{g_{0}} \chi .
\eeqn
At tree level the inverse propagators of the relevant fields,
 as derived from the lagrangian 
(\ref{lag}),  are
\beqn
\begin{array}{ll}
D^{-1}_{ab}=\delta_{ab} (k^{2} +\chi)\, , & \hspace{1cm}
D^{-1}_{\s \s}=k^{2} +\chi \, ,\\
D^{-1}_{\s \tilde{\chi}}=\vf \, , & \hspace{1cm}
D^{-1}_{\tilde{\chi} \tilde{\chi}}=\fr{-1}{g_{0}N}\, ,
\end{array}
\eeqn
and they are drown in fig.~\ref{propa}.

Using eqs.~(\ref{www}), (\ref{renc}) and expanding the potential as in 
eq.~(\ref{mare}), we are now ready to compute
the effective potential at higher orders in the 
$1/N$ expansion.
Now we   show the result of the calculation of
 the first two terms of the  potential expanded as in eq.~(\ref{mare}).

\subsection{Leading order effective potential}

In order  to compute the
effective potential beyond the tree level, we add to $V_{tree}$
the contribution of the bubble diagrams shown in fig.~\ref{cerchi}.a.
 These diagrams give
\beqn
\fr{N-1}{2} \int \ln (k^{2} +\chi) \sim \fr{N}{2} \int \ln (k^{2} +\chi) \, ,
\label{puz}
\eeqn
where we have taken only the term proportional to $N$.
The subleading term $-1/2 \int \ln (k^{2} +\chi)$ contributes to NLO effective
potential $V_{NL}$, see below.

 Using eq.~(\ref{puz}), the unrenormalized LO potential at one loop
 is given  by 

\beqn
N \; V_{L}=-\fr{\chi^{2}}{2 g_{0}}+ \fr{N}{2} \chi \vf^{2}+
\fr{m_{0}^{2}}{g_{0}} \chi +\fr{N}{2}\int \ln (k^{2} + \chi) \, .
\eeqn  

In order to renormalize $V_L$ we impose the renormalization conditions
eq.~(\ref{renc}).
At this order these conditions are given by
\beqn
\nonumber
& & \chi_{0} (\vf^{2})|_{\vf=0}=\mu^{2} \, ,
\\ \nonumber 
& & \left.
\fr{d \, V^{Ren}_{L}(\vf ,\chi(\vf))}{d \,\vf^{2}}
\right|_{\vf=0,\;\chi_0=\mu^{2}}=
\fr{\mu^{2}}{2} \, ,  
\\ 
& & \left.
\fr{d^{2} \, V^{Ren}_{L}(\vf ,\chi(\vf))}{d^{2} \,\vf^{2}}
\right|_{\vf=0,\;\chi_0 =
\mu^{2}}=\fr{g \ N}{4} \, , 
\label{picasso}
\eeqn

The final result for the renormalized
LO potential is readily found
 
\beq
V^{Ren}_{L}=-\fr{\chi_{0}^{2}}{2 g}+ \fr{N}{2} \chi_{0} \vf^{2}+ 
\fr{\mu^{2}}{g}\chi_{0}  \left(1+\fr{g N}{2 (4 \pi)^{2}}\right)+
\fr{N}{4 (4 \pi)^{2}} \chi_{0}^{2} \left( \ln (\fr{\chi_{0}}{\mu^2}) 
-\fr{3}{2} \right) \, ,
\eeq
where $\chi_{0} $  is the solution of eq.~(\ref{czero}) which 
 written in terms of the 
renormalized potential and fields becomes 

\beq
- \fr{\chi_{0}}{ g} +\fr{N}{2}  \vf^{2}+
\fr{\mu^{2}}{g}\left(1+\fr{g N}{2 (4 \pi)^{2}}\right)+
\fr{N}{2 (4 \pi)^{2}} \chi_{0} \left(\ln (\fr{\chi_{0}}{\mu^{2}})-1 \right)=0
\, .
\label{olo}
\eeq

As noticed by \cite{coleman} the effective  potential is afflicted by
pathologies
for large values of $\vf^{2}$.
  The solution of (\ref{olo}) 
increases monotonically in $\vf^{2}$ from 
the value $\chi_{0}=\mu^{2}$ at $\vf^{2}=0$  to
$\chi_{0}=\mu^{2} \exp (32\pi^2/gN) $ at $\vf^{2}=\vf^{2}_{max}$, where 
\beq
\vf^{2}_{max}= \mu^2 \left[
\fr{ -2}{gN}\left(
1+\fr{gN}{32\pi^2}\right) +\fr{\exp\left(\fr{32\pi^2}{gN}\right)}{32\pi^2}
 \right]\, .
\eeq
For $\vf^{2}>\vf^{2}_{max}$ 
 the solution of  equation (\ref{olo})
is found  at
  complex values of $\chi_{0}$.
 Consequently  also $V_{L}$  becomes complex.

\subsection{NLO potential}

\begin{figure}
\centerline{\hbox{\psfig{figure=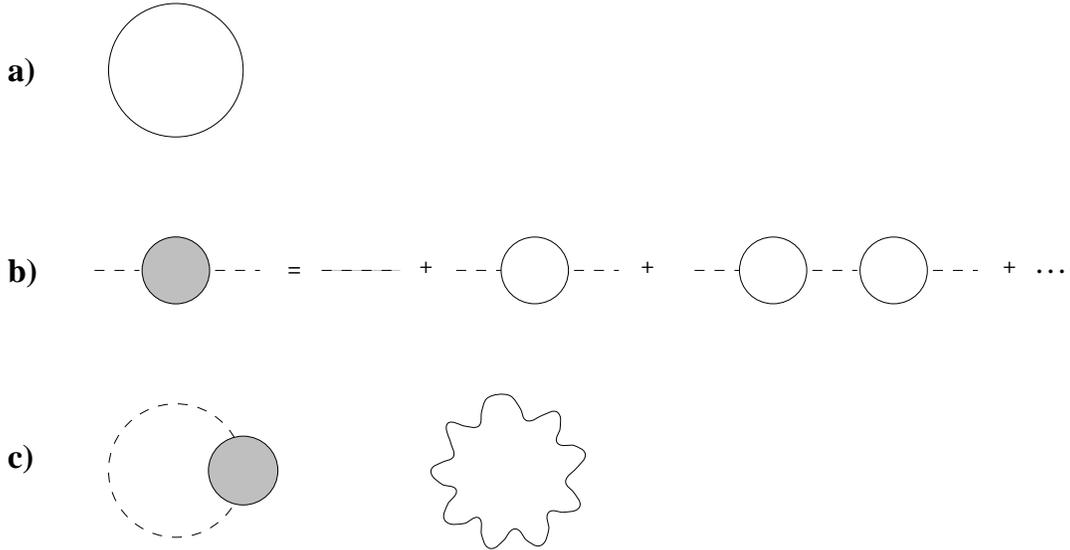,width=14cm,angle=0}}}
\vskip1cm
\caption{Contribution to $V(\vf ,\chi )$  at LO and NLO: (a) the
 LO contribution; (b) the 
propagator of the $\tilde{\chi}$-field at LO, i.e. with all
 $\vf_{a}$-bubbles resummed;
(c) the NLO contributions.}
\label{cerchi}
\end{figure}

In order to compute the NLO effective potential we proceed as follows. 

 a) First we  calculate the propagator of $\tilde{\chi}$    
by resumming 
  all $\vf_{a}$-bubble diagrams, as shown  in fig.~\ref{cerchi}.b .
 The result is

\beqn
D^{-1}_{\tilde{\chi} \tilde{\chi}}=\fr{-1}{gN} (1+gN I(k^{2},\chi ))\, ,
\eeqn
where $I(k^{2},\chi )$ is the renormalized bubble diagram.
In dimensional
regularization we obtain
\beqn
\nonumber
 I(k^{2},\chi )
& =& \fr{1}{2}\left[ \mu^{2 \epsilon}
\int d {\tilde{p}} 
\left( \fr{1}{(p^{2}+\chi )((k+p)^{2} +\chi)} \right) -
\fr{1}{(4\pi)^{2}} \left( \fr{1}{\epsilon}-\gamma_{E}\right)\right]=
\\ 
&=&\fr{1}{2 (4 \pi)^{2}} 
\left[ 2 -\sqrt{1+4 \fr{\chi}{k^{2}}} \ln 
\left( \fr{k^{2} +2 \chi +k \sqrt{k^{2}+4\chi }}{2 \chi} \right)
-\ln \fr{\chi}{\mu^{2}} \right]\, ,
\eeqn
where
$ d \tilde{p}=d^{4-2 \epsilon} p/(2 \pi)^{4-2 \epsilon}$
is the $d$-dimensional phase space and $\epsilon=2-d/2$.

b) We construct with $D^{-1}_{\tilde{\chi} \tilde{\chi}}$ 
the NLO effective  potential from the diagrams shown in fig.~\ref{cerchi}.c.
We have 
\beqn
\fr{1}{2} \ln \left\{ \det \left(
\begin{array}{lr}
k^{2} +\chi & \vf \\
\vf & -\fr{1+g N I(k^{2},\chi)}{g N} \\
\end{array}
\right) \right\} \, .
\eeqn
This term  must  be added
to   the subleading part  of equation (\ref{puz}), as we mentioned before.

The final formula for the  unrenormalized NLO potential is given by

\beqn
V_{NL}=\fr{1}{2} \int \left[-\ln \left(-\fr{gN}{1+gNI}\right) +
 \ln \left( 1+\fr{\vf^{2}}{k^{2} + \chi} \fr{gN}{1+gNI} \right) \right]\, .
\label{v1} 
\eeqn

\subsubsection{Renormalization of $V_{NL}$}

In order to renormalize the potential $V=V_{L}+V_{NL}$ 
we have to impose the renormalization
conditions of eq.~(\ref{renc}).
Since at the lowest order we have imposed the conditions in 
eq.~(\ref{picasso}), the renormalization conditions for $V_{NL}$ are 
given by 
\beqn
& & \chi_{1}(\vf^{2})|_{\vf=0}=0 \, ,
\nonumber
\\
 \nonumber 
& & \left.
\fr{d \, V^{Ren}_{NL}(\vf ,\chi(\vf))}{d \,\vf^{2}}
\right|_{\vf=0,\;\chi_0=\mu^{2}}=0
 \, ,  
\\ 
& &\left.
\fr{d^{2} \, V^{Ren}_{NL}(\vf ,\chi(\vf))}{d^{2} \,\vf^{2}}
\right|_{\vf=0,\;\chi_0 =
\mu^{2}}=0 \, , 
\label{renoir}
\eeqn

The renormalization procedure is complicated by the presence of 
quadratic 
divergences in some of the integrals. The details of the calculations
 are given in  appendix A.
Here we give the final result:

\beqn
\label{vnl}
& &V_{NL}^{Ren}= V_{1}+V_{2}\, ,
\eeqn
\beqn
\nonumber
& & V_{1}=\fr{1}{2}
\int \ln \left( 1+\fr{\vf^{2}}{k^{2} + \chi} \Delta \right) -
\fr{\vf^{2}}{2}\int \fr{\Delta}{k^{2}+\mu^{2}}
+\fr{\vf^{4}}{4}  \int  
  \fr{\Delta^{2}}{(k^{2}+\mu^{2} )^{2}}  +
\\
& &+ \fr{\vf^{2}}{2} (\chi -\mu^{2})\int 
\fr{\Delta}{k^{2}+\mu^{2}} \left(
 \fr{1}{k^{2}+\mu^{2}} +
I'(k^{2},\mu^{2})\, \Delta
\right) 
+ \fr{\vf^{2}}{2} \; f(\chi )
 \fr{1}{(4 \pi)^{2}} \int 
\fr{\Delta^{2}}{(k^{2}+\mu^{2})^{2}} \, ,
\label{babbo} 
\eeqn
\beqn
 \nonumber
  & &\!\!\! V_{2}=
\fr{-1}{2} \! \int \ln \left(-\Delta \right)+
\fr{(\chi -\mu^{2})}{2} \! \int \Delta \; I'(k^{2},\mu^{2}) 
-\fr{(\chi -\mu^{2})^{2} }{4}\!
\int \Delta
\left(
 I''(k^{2},\mu^{2})-\Delta \; I'^{2}(k^{2},\mu^{2})
\right) +
\\
\nonumber
& & +\fr{3}{2(4\pi)^{2}} g(\chi)\!\!
\int
 \Delta^{2} 
\fr{ I'(k^{2},\mu^{2})}{k^{2}+\mu^{2}}+
\fr{3}{2(4\pi)^{2}} h(\chi) \!
\int   \!
\fr{ \Delta^{2}}{(k^{2}+\mu^{2})^{2}}
+ \fr{l(\chi )}{2(4 \pi)^{2}} \!  \int
\fr{\Delta}{k^{2}+\mu^{2}}
\left( 1+\fr{\mu^{2}}{k^{2}+\mu^{2}}\right) \! \! + 
\\
& &+\fr{ g(\chi)}{(4 \pi)^{2}}  \; \int
 \fr{\Delta}{(k^{2}+\mu^{2})^{2}} 
-\fr{h(\chi)}{2 (4 \pi)^2} \int
\fr{\Delta^{2}}{k^{2}+\mu^{2}} I'(k^{2},\mu^{2})-
\fr{j(\chi) }{4 (4 \pi)^{4}}\int
\fr{\Delta^{2}}{(k^{2}+\mu^{2})^{2}}\, ,
\label{mamma} 
\eeqn
\vspace{.5cm}

\noindent
where
 $\Delta = gN /[1+gN I(k^2,\mu^{2} )] $ and
 the exact form of $f(\chi)$, $g(\chi)$, $h(\chi)$, $l(\chi)$, $j(\chi)$
is given in the appendix A.

The pathologies of 
 $V_{NL}$  look even  more serious than in the LO case. 
As  noticed in \cite{root},  $V_{NL}$ is complex at every value
of $\vf$.
The reason is that 
$D_{\tilde{\chi} \tilde{\chi}} (k^{2},\chi) $ crosses the (Landau) pole,
$\Lambda_{Landau}^{2}=\mu^2  \exp{(2+32 \pi^{2}/Ng)}$,  
for large values of $k^{2}$, so
that the integration over $k^{2}$ always leads to a complex result.
The analytic properties of the potential are better understood
 in the framework of the renormalon theory.
In order to isolate the contribution of the different renormalons it is better
to work in the space of the Borel trasform.

\section{Borel trasform of $V_{NL}$}

In this  section we perform  the Borel trasform of the effective potential.
The essential ingredients  of this calculations are the Borel trasform of
$-gN/(1+gNI)$ and its powers:

\beq
B\left[\fr{-gN}{1+gNI} \right]= N \exp{(-N t I)}\, ,
\hspace{1.5cm}
B\left[\fr{-gN}{1+gNI} \right]^{n}= N \fr{(N t)^{(n-1)}}{(n-1)!} \exp{(-N t I)}
\, ,
\label{b2}
\eeq
where $t$ is the Borel parameter conjugate of the coupling costant $g$ of 
eq.~(\ref{borel}).
It is important to note that even though 
$I=I(k^{2}, \chi)$ is a funtion of the constraint
$\chi$,  and  depends on the coupling costant $g$ via eq.~(\ref{olo}),
we can perform the Borel trasform  before imposing the constraint 
 eq.~(\ref{olo}).
In this way we can treat $\chi$ as a variable indipendent of $g$.

The Borel trasform of  the different terms  appearing  in $V_{NL}^{Ren}$ can 
be obtained in
 the following way.

i) For the first term  in eq.~(\ref{mamma}) we have:

\beqn
\nonumber
& &B\left[ \ln \left( -  \fr{gN}{1+gNI}  \right) \right]=
\left. \fr{d}{d h} 
 \right|_{h=0} B\left[ \left( -  \fr{gN}{1+gNI}  \right)^{h} \right]=
\\ 
& & =\left. \fr{d}{d h}  \right|_{h=0} 
 \fr{(N t)^{h-1}}{ \Gamma (h)}  B\left[ \left( -
\fr{gN}{1+gNI}  \right) \right] =
\fr{1}{N t} B\left[ \left( -
\fr{gN}{1+gNI}  \right) \right]\, .
\eeqn

ii) The first term in eq.~(\ref{babbo}) is  given by
\beqn
\nonumber
& &B\left[ \ln \left( 1+\fr{\vf ^{2}}{k^{2} + \chi} \fr{gN}{1+gN}
\right) \right]
=B\left[ \sum _{n=1} \fr{(-)^{n+1}}{n} \left(
\fr{\vf ^{2}}{k^{2} + \chi}\right)^{n}
\left(\fr{gN}{1+gNI}\right)^{n}\right] =
\\
& &=- \fr{1}{Nt} B\left[ -\fr{gN}{1+gNI}\right] \sum_{n=1} \fr{1}{n!} \left(
\fr{Nt \vf ^{2}}{k^{2} + \chi} \right)^{n}=
\fr{1-\exp \left( \fr{Nt \vf ^{2}}{k^{2} + \chi} \right)}{Nt} 
\; B \; \left[-\fr{gN}{1+gNI}\right] \, .
\eeqn

iii) The Borel trasform of all the other terms can be be done straightforwardly
substituting $\Delta$,  $\Delta^{2}$
respectively with $- N \exp(-Nt \, I)$ and $N^2 t \,\exp(-Nt \, I)$, 
according to eq.~(\ref{b2}).

From i) and ii) we get
the final expression of the Borel trasform of unrenormalized
$V_{NL}$ 

\beq
\fr{-1}{2 t} \int
 \exp \left[ N t\left( \fr{\vf^{2}}{k^{2}+\chi} -I(k^{2},\chi)
\right) \right]\, ,
\label{fagiolo}
\eeq
and iii) is used in the  whole renormalized potential,
that is $V_{NL}^{Ren}$.
Notice that, if in eq.~(\ref{fagiolo}) we consider  only the asymptotic
behaviour of $I(k^{2},\chi)$, 
\beq
\exp(-N t\, I) \sim  e^{-2 u}
 \left(\fr{k^{2}}{\mu^{2}} \right)^{u} \;\;\;\;\;\; \hspace{2cm}
 u=\fr{Nt}{32 \pi^{2}}\, ,
\label{uuu}
\eeq
then some   renormalon contributions are missed  in $V_{NL}$.

On the other hand, since the exact calculation of the Borel trasform 
is very hard, some
approximation   is necessary. This is
the argument of the next  section.

\section{First renormalon in the NLO effective potential}

Using the Borel trasform 
of eq.~(\ref{vnl}), it is possible to isolate the contribution of
the first renormalon to the potential
and to  write  the corresponding  renormalon contribution
to the  1-PI green functions.
Our strategy is the following:
\begin{enumerate}
\item[a)]
we expand the Borel trasform of $V^{Ren}_{NL}(\vf ,\chi)$
 in powers of $\chi$ and $\vf$ around the point $\vf=0$, $\chi=\mu^{2}$;
 
\item[b)] in this expansion we encounter integrals of the form 
\beqn
\int
\fr{ [I^{(a_{1})}(k^{2},\mu^{2})]^{b_{1}} \, 
 [I^{(a_{2})}(k^{2},\mu^{2})]^{b_{2}} \, ...}{(k^{2}+\mu^{2})^{c}}
\exp[-Nt \, I(k^{2},\mu^{2})]\, ,
\eeqn 
where $I^{(a)}(k^{2},\mu^{2})=\left. [\partial^a 
I(k^{2},\chi)/\partial\chi^a] \right|_{\chi=\mu^{2}}$.
 Renormalon contributions come from the UV behaviour of such integrals, 
the integrands of which at large momenta typically behave as 
$
\ln^{m}(k^2)/(k^2)^{n-u}
$,
where $u =Nt/(32\pi^2)$ is the Borel variable.
In dimensional regularization, $u$ acts as a regulator and the above
UV behaviour yields poles
for $u=n-2$. For the unrenomalized NLO potential,
 $n\geq 1$, so  that poles develop at $u=-1,0,1,...$ .  
The singularities in $u=-1,0$ correspond to the quadratic and 
the logarithmic
divergencies of the potential and are removed by renormalization.
 Thus in the renormalized NLO potential, the pole closest to the origin
 is located at $u=1$;

\item[iii)] we calculate 
 the first renormalon contribution by
  considering all integrands whose ultraviolet behaviour
is  $\ln^{m} (k^{2})\,  k^{(u-3)}$.
Let us note that the counterterms added in order to renormalize $V_{NL}$ not 
 only remove the poles for $u=-1,0$, but give also contributions to the 
renormalons at $u=1,2,3,...$. Thus the counterterms are
fundamental to obtain the final result.
\end{enumerate}

We skip all the details of this complicated
calculation and we report only the final result.
Denoting $\tilde{V_{1}}$ and $\tilde{V_{2}}$
as the first renormalon contributions to $V_{1}$ and $V_{2}$,
eqs.~(\ref{babbo}) and (\ref{mamma}),
we have

\beqn
\nonumber
& & \tilde{V_{1}}=
\fr{\vf^{6} (Nt)^{2}}{12(1-u)}+
\fr{\vf^{4} }{2}
\left[ 
\fr{N t}{1-u}(\chi-\mu^{2})+ \fr{(N t)^{2}}{2 (4\pi)^{2}} \left(
\fr{\mu^{2}-\chi}{(1-u)^{2}}+\fr{f(\chi)}{1-u}\right) \right] + 
\\
\nonumber
& &+ \fr{\vf^{2}(\chi-\mu^{2})^{2}}{2}
\left(\fr{1}{1-u}-\fr{2 (Nt)}{(4\pi)^{2}(1-u)^{2}}+ 
\fr{ (Nt)^{2}}{(4\pi)^{4}(1-u)^{3}}\right) +
\\
\nonumber
& &+ \fr{\vf^{2} (Nt)}{2(4\pi)^{2}}
\left[ 
\fr{1}{1-u}
 \left(
 2\chi^{2}-3\mu^{2} \chi +\mu^{4}-\mu^{2} \chi
\ln  \fr{\chi}{\mu^{2} }  \right)+
\left(
\fr{ 5}{(1-u)} - 
\fr{3 (Nt)}{(4\pi)^{2}(1-u)^{2} }
\right) g(\chi)
\right] +
\\
& &+ \fr{3\vf^{2} (Nt)^{2}}{2 (4\pi)^{2} (1-u)}
\left[ 
\fr{7\chi^{2}}{8}-\mu^{2} \chi+\fr{\mu^{4}}{8}+
\fr{\chi^{2}}{4} 
\left(
\ln^{2}  \fr{\chi}{\mu^{2}} -3\ln 
 \fr{\chi}{\mu^{2}}\right) \right]\, .  
\label{r1}
\eeqn
For $\tilde{V}_2$
it is more convenient to present its third derivative 
$\partial^3 \tilde{V}_2/\partial x^3$  respect to $\chi$. 
Thus, $\tilde{V_{2}}$ can easily be obtained integrating three times in 
$\chi$ with the conditions 
$\tilde{V}_{2}(\mu^2)=\tilde{V}'_{2}(\mu^2)=\tilde{V}''_{2}(\mu^2)=0$,
\beqn
\nonumber
& & \fr{\partial^{3} \tilde{V}_{2}}{\partial \chi^{3}}=
-\fr{3 (Nt)^{2}}{(4 \pi)^{6}(1-u)^{4}}+
\fr{1}{(4 \pi)^{4}(1-u)^{3}}\left(
6 (Nt)+\fr{(Nt)^{2}}{(4 \pi)^{2}} \left(
1-\fr{\chi}{\mu^{2}}+7 \ln \fr{\chi}{\mu^{2}}\right) \right)+
\\ 
\nonumber
& &+\fr{1}{(4 \pi)^{2}(1-u)^{2}} \left[
-6+\fr{(Nt)}{(4 \pi)^{2}}\left( -11\ln\fr{\chi}{\mu^{2}}+2\fr{\mu^{2}}{\chi}+
\fr{\chi}{\mu^{2}}-\fr{15}{2}\right)+ \right.
\\
\nonumber
& &
\left.
+\fr{(Nt)^{2}}{(4 \pi)^{4}} \left(
-\fr{7}{2}\ln^{2} \fr{\chi}{\mu^{2}}
+\fr{3}{4}\ln\fr{\chi}{\mu^{2}}+\fr{7}{8}-\fr{\chi}{\mu^{2}}
+\fr{\chi^{2}}{8\mu^{4}}\right) \right]+
\fr{1}{(4 \pi)^{2}(1-u)} \left[
6\ln\fr{\chi}{\mu^{2}}+16-2\fr{\mu^{2}}{\chi}+\right.
\\
\nonumber
& &
\left.
-\fr{\mu^{4}}{2\chi^{2}}
+\fr{(Nt)}{(4 \pi)^{2}}\left(
 4\ln^{2}\fr{\chi}{\mu^{2}}+14\ln\fr{\chi}{\mu^{2}}-
\fr{\mu^{2}}{2\chi}\ln\fr{\chi}{\mu^{2}}
-\fr{3\chi}{2\mu^{2}}+\fr{3\mu^{2}}{\chi}+\fr{9}{2}\right)+
\right.
\\
& &
\left.
+\fr{(Nt)^{2}}{(4 \pi)^{4}} \left(
\fr{7}{6} \ln^{3} \fr{\chi}{\mu^{2}}
-\fr{3}{8} \ln^{2} \fr{\chi}{\mu^{2}}
-\fr{7}{8} \ln \fr{\chi}{\mu^{2}}
-\fr{\chi^{2}}{16\mu^{4}}
+ \fr{\chi}{\mu^{2}}-\fr{15}{16}\right) \right]\, .
\label{r2}
\eeqn

In the previous formulae (\ref{r1}) and (\ref{r2})  we have omitted an overall
 factor $e^{-2 u}/((4\pi)^{2}\mu^{2})$,
which is, however,  essential to the calculation of the residue at $u=1$.

 The formulae (\ref{r1}) and  ( \ref{r2}) represent the main result of 
this paper and some comments are  necessary for a better
understandig.

First of all we recall  that by applying $n$  total derivatives
\beq
\fr{d^n}{(d \vf)^n}=
\left( \fr{\partial }{\partial \vf}+\fr{d \chi }{d \vf}
\, \fr{\partial }{\partial\chi }\right)^n \, ,
\label{td}
\eeq
computed 
at $\vf=0$, $\chi=\mu^2$ to the effective potential we obtain all
 zero momentum 1-PI Green functions $\Gamma^{(n)}(p=0,g)$.
Thus, acting by  eq.~(\ref{td}) on $\tilde{V}=\tilde{V}_{1}+\tilde{V}_{2}$ we 
compute  the first renormalon contribution to all 1-PI Green functions.
We  see that the	
 first renormalon appears for the first time in  the 6-points 1-PI Green
function. Actually we have
$\left. d\tilde{V}/d\vf^{2} 
\right|_{\vf=0}=\left. d^{2}\tilde{V}/d^{2}\vf^{2} \right|_{\vf=0}=0$ and
\beq
\left.
\fr{d^{3} \tilde{V}}{d (\vf^{2})^{3}} \right|_{\vf=0 }
=
( \chi_{0}')^{3} \fr{\partial^{3} \tilde{V}}{\partial \chi^{3}}+
3 (\chi_{0}')^{2} 
\fr{\partial^{3} \tilde{V}}{\partial \chi^{2}\partial \vf^{2} }+
3 \chi_{0}' \fr{\partial^{3} \tilde{V}}{\partial \chi \partial \vf^{4}}+
\fr{\partial^{3} \tilde{V}}{\partial \vf^{6} }\, ,
\label{g6}
\eeq
where, from eq.~(\ref{olo})
\beq
\left.
\chi_{0}'=\fr{d\chi}{d \vf^2} \right|_{\vf=0}=\fr{g N}{2}\, . 
\eeq
In eq.~(\ref{g6}) the residue to the pole in $u=1$  give us
\beqn
\nonumber
Im \Gamma^{(6)}(p=0,g)
& =& -\fr{375}{16\pi \mu^{2}}e^{-2} g^{3} e^{-32\pi^{2}/(g N)}
+O( e^{-64\pi^{2}/(g N)})=
\\
&=&-\fr{375}{16\pi \Lambda_{Landau}^{2}} g^{3} 
+O(\fr{1}{ \Lambda_{Landau}^{4}})\, .
\label{lan}
\eeqn  
 The $Im \Gamma^{(6)}(p=0,g)$ has been also  calculated by \cite{dicecio}
summing all  Feynman graphs that contribute to the first renormalon of the
$\Gamma^{(6)}$. We can easily obtain this diagrammatic expansion replacing
 $\tilde{V}$ with $V_{NL}^{Ren}$, eq.~(\ref{vnl}),  in  eq.~(\ref{g6}) as
 shown in appendix B.
From eq.~(\ref{babbo}) and (\ref{mamma}) it follows also that, in general
$\left. d^{n}\tilde{V}/d^{n}\vf^{2} \right|_{\vf=0}\neq 0$ for $n\geq 3$ due
to the presence of terms like $\ln \chi/\mu^{2}$ and
$1/\chi$, i.e. 
  the first renormalon contribution is present in 
 any $2n$-point Green function  with $n\geq 3$.

\begin{figure}
\centerline{\hbox{\psfig{figure=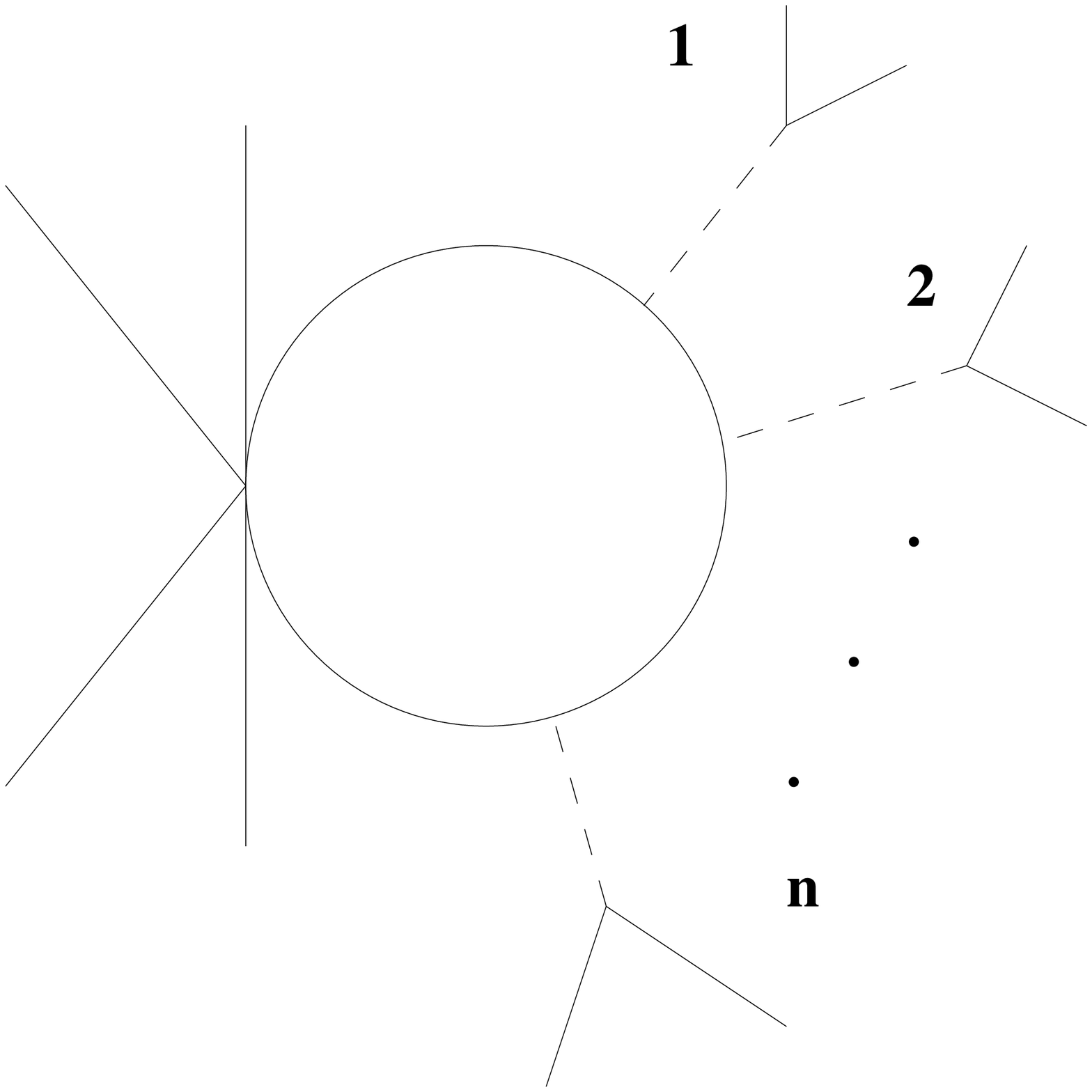,width=10cm,angle=0}}}
\vskip1cm
\caption{$2(n+2)$-point 1-PI Green functions with the insertion of 
$(\vec{\phi}^2)^{3} $ at the leading order.}
\label{gamman}
\end{figure}

Parisi hypothesis  tell us that 
the first renormalon in the Green function can be removed with the insertion of
all possible 6-dimension
operators.
Considering only the zero momentum Green function   
the whole set of such operators reduces to $(\vec{\phi}^2)^3$.
By adding to the Lagrangian the operator $(\vec{\phi}^2)^3$
with a suitable coefficient the first renormalon contribution
 on   all zero momentum Green functions can be removed.
We  have \cite{parisi}
\beq
Im\Gamma^{(n)}(p=0 ,g)=C_{6}(g)\Gamma^{(n)}_{\vf^{6}}(p=0,g)+
O(e^{-64\pi^2/(g N)})
\label{imagine}
\eeq
where $\Gamma^{(n)}_{(\vec{\phi}^2)^3}(p=0,g)$ is the
 n-point 1-PI Green function
with the insertion of $(\vec{\phi}^2)^3$ at zero momentum calculated at
 the leading order in $1/N$  expansion,
whose  diagram is shown in 
fig.~\ref{gamman}.
 At leading order $\Gamma^{(6)}_{(\vec{\phi}^2)^3}(p=0,g)=1$,
and  so $C_6(g)$ is given by eq~(\ref{lan}).
We notice that
  renormalization group analysis \cite{parisi} tells us only that $C_6(g)$ is 
proportional to $g^{3} e^{-32\pi^{2}/(g N)}$. By
performing derivatives on $\tilde{V}$ we can verify that eq.~(\ref{imagine})
holds for every $n$. For example
we have explicitly checked that 
$Im \Gamma^{(8)}(p=0,g)$ satisfies  eq.~(\ref{imagine}).

\section{Conclusions}

One of the main features that emerges from the Borel trasform of 
$V_{NL}$ is the structure of the  singularities in the 
Borel parameter $u$, due to the UV behaviour 
of Green functions.
We have checked  that for the 
unrenormalized potential these singularities come for integer values of $u$
starting from $u=-1$. 
The renormalization procedure removes the poles at $u=-1,0$.
More precisely the insertion of operators of dimension 2 remove the
singularity in $u=-1$ (mass  renormalization),
 the operators of dimension four the one in $u=0$
 (coupling costant and wave function renormalization).
In the renormalized potential the singularities are then only found
starting from $u=1$. 
In our study we have shown that indeed the pole in $u=1$ is canceled
by adding a term proportional to the operator 
$(\vec{\phi }^{2} )^{3}$ to the Lagrangian.  The poles in $u=2,3$, etc.
are removed by the insertion of even higher dimension operators.
The resulting theory thus requires  fixing the  renormalization
conditions  for all possible $2n-$point Green functions
(including those containing derivatives) and this would introduce
an infinite number of arbitrary parameters.
In a ``complete" theory we would expect that renormalon singularities do not
appear and that all Green functions are determined in terms 
of only a finite number of parameters.
For ``complete" we mean a theory which appears as an
effective  $(\vec{\phi}^2)^2$-theory
below same physical scale that we call $\Lambda_{Landau}$.

 We have seen  that renormalon effects are suppressed by powers of $\Lambda_{Landau}$.
The first renormalon give a contribution proportional to 
$\Lambda_{Landau}^{-2}$, the second one a contribution proportional 
to $\Lambda_{Landau}^{-4}$ and so on.
The ambiguities  become then more and more  important as  we move closer to the
region of  the Landau pole. The
renormalon analysis gives us a systematic procedure to order 
the effects of the presence of the Landau pole (scale of new physics) in inverse
powers of $\Lambda_{Landau}$.

\section{Acknowledgments}
 
Both of us want  to thank  sincerily Prof. G. Martinelli for enlightening 
discussions and for his patience in reviewing the manuscripts of this paper.

\section{Appendix A}

In this appendix  we describe in detail the renormlization of the 
effective potential. Let us start with
\beqn
V_{1}=\fr{1}{2} \int 
\ln
\left( 1+\fr{\vf^{2}}{k^{2} + \chi } \fr{gN}{1+gNI(k^2 ,\chi)} \right)\, .
\eeqn
In order to renormalize $V_{1}$  we calculate the two and four point 1-PI
 functions at the point $\vf=0$ and $\chi=\mu^2$.
In the following we put $\Delta(k^2, \chi)= g N/[1+gN I(k^2 ,\chi)]$.
The $V_1$ contribution to $\Gamma^{(2)}$ is
\beqn
\fr{\partial V_{1}}{\partial \vf^{2}}=
\fr{1}{2} \int
\fr{1}{k^{2}+\chi} 
\Delta(k^2, \chi)\, .
\eeqn
In oder to isolate the divergences
we develop $\partial V_{1}/\partial \vf^{2}$
 around  $\chi=\mu^{2}$,
\beqn
\nonumber
\fr{\partial V_{1}}{\partial \vf^{2}} &=&
\fr{1}{2} \int 
\fr{1}{k^{2}+\mu^{2}}
\Delta(k^2,\mu^2)+ 
\\ \nonumber 
&-&\fr{1}{2} (\chi -\mu^{2})\int 
\fr{1}{k^{2}+\mu^{2}} 
\Delta(k^2,\mu^2) 
\left( \fr{1}{k^{2}+\mu^{2}} +
I'(k^{2},\mu^{2})
\Delta(k^2,\mu^2)
\right) +
\\ \nonumber  
&-& \fr{1}{4} (\chi -\mu^{2})^{2} \int 
\fr{I''(k^{2},\mu^{2})}{k^{2}+\mu^{2}}
\Delta^{2}(k^2,\mu^2)+
\\ \nonumber  
&+& \fr{1}{2}(\chi -\mu^{2})^{2} \int 
\fr{1}{(k^{2}+\mu^{2})^{2}}
\Delta(k^2,\mu^2) 
\left(
\fr{1}{k^{2}+\mu^{2}} +
I'(k^{2},\mu^{2})
\Delta(k^2,\mu^2) 
\right) + 
\\ 
&+&\fr{1}{2}(\chi -\mu^{2})^{2} \int 
\Delta^{3}(k^2,\mu^2)
(I')^{2}(k^{2},\mu^{2})
\fr{1}{k^{2}+\mu^{2}}
 + \, ... \; .
\label{cac}
\eeqn
At large momenta $I(k^2) \sim \ln (k^{2})\,\, $ and  
$I' ,I'' ,..\sim 1/k^{2}$.
Thus the divergent parts of the previous equation are
all contained by the terms in the first three
lines of eq.~(\ref{cac}).
The divergencies present in the  first two terms
can be removed by adding as counterterm 
\beq
-\fr{\vf^{2}}{2} \int \left[
\fr{1}{k^{2}+\mu^{2}}
\Delta(k^2,\mu^2)
+ (\chi -\mu^{2})
\fr{1}{k^{2}+\mu^{2}}
\Delta(k^2,\mu^2)
\left( \fr{1}{k^{2}+\mu^{2}} +
I'(k^{2},\mu^{2})
\Delta(k^2,\mu^2)
\right) \right]\, .
\eeq
The leading, large momentum behaviour of $I''$ is
\beqn
I''(k^{2} ,\chi )=\fr{1}{(4 \pi)^{2}} \left[
\fr{1}{\chi (k^{2} +4\chi)}  
+O(\fr{1}{k^{4}})
\right]\, .
\eeqn
The fact that  $I''(k^{2} ,\chi )-1/[(4 \pi)^{2}\ \chi \ 
(k^{2}+\mu^{2})]=O(1/k^{4})$  suggests
that the missing counterterm has of the following form
\beqn
\fr{1}{2} \vf^{2} f(\chi ) \int \fr{1}{(4 \pi)^{2}} 
\fr{1}{(k^{2}+\mu^{2})^{2}}
\Delta^{2}(k^2,\mu^2)\, , 
\eeqn
where $f(\chi)$ satisfies 
$f(\mu^{2})=f'(\mu^{2})=0$ and $f''(\chi)=1/\chi $, 
\beqn
f(\chi)=\mu^{2}-\chi +\chi \ln \fr{\chi}{\mu^{2}} \, .
\eeqn
The only non-zero contribution of $V_1$ to  $\Gamma^{(4)}$ for $\vf =0$ and
$\chi =\mu^{2} $, is given by
\beqn
\fr{\partial^{2} V_{1}}{\partial \vf^{4}}=
-\fr{1}{2} \int \left(
\Delta(k^2,\chi)\,
\fr{1}{k^{2}+\chi} \right)^{2}\, .
\eeqn
This can be cancelled by  with the following counterterm
\beq
\fr{\vf^{4}}{4} \int \left(
\Delta(k^2,\chi)\,
\fr{1}{k^{2}+\mu^{2}} \right)^{2}\, .
\eeq
We can apply a similar procedure to
\beq
V_{2}= - \fr{1}{2} \int \ln (- \Delta(k^2,\chi) )\, .
\eeq
In this case there is no  explicit dependence on $\vf^{2}$,
and the expansion of  $V_2$ around $\chi =\mu^{2}$ gives
\beqn
\nonumber
V_{2}&=&\fr{1}{2} \int \left[ -\ln (- \Delta(k^2,\mu^2) )
+ (\chi -\mu^{2}) \Delta(k^2,\mu^2) 
I'(k^{2},\mu^{2}) \right] + 
\\ 
\nonumber
&+& \fr{(\chi -\mu^{2})^{2} }{2}
\Delta(k^2,\mu^2)
\left[
 I''(k^{2},\mu^{2})-\Delta(k^2,\mu^2) I'^{2}(k^{2},\mu^{2})
\right]+
 \\
 \nonumber  
&+&\fr{(\chi -\mu^{2})^{3}}{6}
\Delta(k^2,\mu^2)
\left[
I'''(k^{2},\mu^{2})-
3 \Delta(k^2,\mu^2) I'(k^{2},\mu^{2})I''(k^{2},\mu^{2}) 
\right]+
\\
 \nonumber
&+&\fr{(\chi -\mu^{2})^{3}}{3}
\Delta(k^2,\mu^2)^3
I'^{3}(k^{2},\mu^{2})
  +\, ...\; .
\eeqn

The divergences of the first two lines can be simply substracted.
The term
\beq
-3 
\Delta^{2}(k^2,\mu^2)
 I'(k^{2},\mu^{2})I''(k^{2},\mu^{2})\, ,
\label{azz}
\eeq
is  logaritmically  divergent. It can be kept  finite  adding to $V_2$
\beq
\fr{1}{2}\int
\fr{3}{(4 \pi)^{2}} 
\Delta^{2}(k^2,\mu^2)
\fr{ I'(k^{2},\mu^{2})}{k^{2}+\mu^{2}} g(\chi)\, ,
\eeq
with $ g(\chi)$ such that
$g(\mu^{2})=g'(\mu^{2})=g''(\mu^{2})=0$,
$g'''(\chi)=1/\chi$.
Thus
\beqn
g(\chi )=\fr{\chi^{2}}{2} \ln (\fr{\chi}{\mu^{2}} )
-\fr{3}{4} \chi^{2} +\mu^{2} \chi-\fr{\mu^{4}}{4}\, .
\eeqn
A further derivative applied to eq.~(\ref{azz})  shows that there is
 a further divergence due to 
\beq
-3 \Delta^{2}(k^2,\mu^2) 
I''^{\,\, 2}(k^{2},\mu^{2}) \, .
\eeq
This can be eliminated by
\beqn
\fr{1}{2}\int
\fr{3}{(4 \pi)^{2}}
\Delta(k^2,\mu^2)^{2}
\fr{ I''(k^{2},\mu^{2})}{k^{2}+\mu^{2}} h(\chi)\, ,
\eeqn
with  $h(\mu^{2})$ and its  first three derivatives  calculated in 
$\chi=\mu^{2}$ equal to zero and
$h^{iv}(\chi)=1/\chi^2$,
\beq
h(\chi)=\fr{\chi^{3}}{6 \mu^{2}}+
\fr{\chi^{2}}{4} - \fr{\mu^{2} \chi}{2} +\fr{\mu^4}{12}-\chi^{2}
\fr{\chi^{2}}{2}
 \ln \fr{\chi}{\mu^{2}}\, .
\eeq
Finally  we have the term
\beq
\Delta(k^2,\mu^2) I'''(k^{2},\mu^{2})\, .
\label{i3}
\eeq
which at  large momenta behaves as
\beq
I'''(k^{2},\chi)=
-\fr{1}{(4 \pi)^{2}}
\left[
\fr{1}{\chi^{2}( k^2 + 4 \chi)}+\fr{6}{\chi ( k^2 + 4 \chi)^2}
+ O \left( \fr{1}{k^6} \right) \right] \sim 
-\fr{1}{(4 \pi)^{2}} 
\left( \fr{1}{\chi^{2} k^{2}}+
\fr{2}{\chi k^{4}} \right)\, .
\eeq
We can renormalize this contribution by adding as counterterm
\beqn 
\fr{1}{2}\int
\fr{1}{(4 \pi)^{2}}
\Delta(k^2, \mu^2)
 \left[
\fr{1}{k^{2}+\mu^{2}}
\left( 1+\fr{\mu^{2}}{k^{2}+\mu^{2}}\right)
l(\chi)
+
\fr{2}{(k^{2}+\mu^{2})^{2}} 
g(\chi) \right]\, ,
\eeqn
where $ l( \chi) = -\chi \ln \left(\chi/\mu^{2}\right)
+\chi^{2}/(2\mu^{2}) -\mu^{2}/2 $.

The residual divergencies are eliminated adding to the potential

\beq
-\fr{h(\chi)}{2 (4\pi)^{2}}\int 
\fr{\Delta^{2}(k^2, \mu^2)}{k^2+ \mu^2}I'(k^2, \mu^2)-
\fr{j(\chi)}{4 (4\pi)^{4}}\int
\fr{\Delta^{2}(k^2, \mu^2)}{(k^2+ \mu^2)^{2}}\, ,
\eeq
where $j(\chi )$ is such that $j(\mu^{2})$ and its first four derivatives
calculated in $\chi=\mu^2$ are zero and
 $j^{(v)}(\chi)=1/\chi^{3}$, 
\beq
j(\chi)=h(\chi)+\fr{\chi^{4}}{24 \mu^{4}}-
\fr{\chi^{3}}{6 \mu^{2}}+\fr{\chi^{2}}{4}-
\fr{\mu^{2} \chi }{6}+\fr{\mu^{4}}{24}\, .
\eeq
All these counterterm  are included in eq.~(\ref{babbo},\ref{mamma}),
 section II.

\section{Appendix B}

In this appendix we give all the terms which contribute to first
renormalon  in  the six points 1-PI Green function $\Gamma^{(6)}(p=0,g)$.
We  compare our result, obtained by 
using the effective potential, with the one of ref.~\cite{dicecio} obtained
in a diagrammatic way.

Replacing $\tilde{V}$ with $V_{NL}^{Ren}$ in eq.~(\ref{g6}),
$\Gamma^{(6)}(p=0,g)$ is given by
\beqn
\Gamma^{(6)}(p = 0,g) = 120 \left.
\left[ ( \chi'_0)^3 \fr{\partial^3 V_2}{\partial
\chi^3} + 3 \chi'_0 \fr{\partial^3 V_1}{\partial \chi \partial \vf^4}+
\fr{\partial^3 V_1}{\partial \vf^6} + 3 (\chi'_0)^2 \fr{\partial^3 V_1}{
\partial \chi^2 \partial \vf^2} \right] \right|_{\vf=0,\chi=\mu^2}\, .
\eeqn
We can write the six points 1-PI function as the sum of the several terms

\beq
\Gamma^{(6)} = \Gamma^{(6)}_a+\Gamma^{(6)}_b+\Gamma^{(6)}_c+\Gamma^{(6)}_d
\, ,
\eeq
where  
\beq
\Gamma^{(6)}_a= 120 \fr{\partial^3 V_1}{\partial \vf^6} =
-120 \int 
 \fr{\Delta^3 (k^2,\mu^2)}{(k^2+\mu^2)^3}\, ,
\eeq
\beq
\Gamma^{(6)}_b = 360 \chi'_0\fr{\partial^3 V_1}{\partial \chi \partial \vf^4} =
-180 g \int
 \left[ \fr{\Delta^2 (k^2,\mu^2)}{(k^2+\mu^2)^3} +
\fr{\Delta^3 (k^2,\mu^2)}{(k^2+\mu^2)^2}I'(k^2, \mu^2) \right]\, ,
\eeq
\beqn
\nonumber
\Gamma^{(6)}_c &=& 360 (\chi'_0)^2 \fr{\partial^3 V_1}{\partial \chi^2 \partial 
\vf^2} = -45 g^2 \int 
\left\{
 \left[ 2\fr{\Delta (k^2,\mu^2)}
{(k^2+\mu^2)^3} - \fr{\Delta^2 (k^2,\mu^2)}{(k^2+\mu^2)}I''(k^2, \mu^2)
  \right]+\right.
\\ 
&+& \left. \left[ 2\fr{\Delta^3 (k^2,\mu^2)}{(k^2+\mu^2)}I'^{2}(k^2, \mu^2) +
\fr{\Delta^2
(k^2,\mu^2)}{(k^2+\mu^2)^2}I'(k^2, \mu^2)
 + \fr{1}{(4 \pi)^2 \mu^2} \fr{\Delta^2
(k^2,\mu^2)}{(k^2+\mu^2)^2} \right] \right\}\, ,
\eeqn
\beqn
\nonumber
\Gamma^{(6)}_d &=& 120 (\chi'_0)^3 \fr{\partial^3 V_2}{\partial \chi^3} =
-15 g^3 \int 
\left\{  \left[ \fr{3}{2} \fr{1}{(4 \pi)^2 \mu^2} \fr{\Delta 
(k^2,\mu^2)}{(k^2+\mu^2)^2} + \Delta^2 (k^2,\mu^2)
 I'^{3}(k^2, \mu^2)  \right]+\right.
 \\ 
\nonumber
&+& \left[ \fr{3}{2} \fr{1}{(4 \pi)^2 \mu^2} \fr{\Delta^2 (k^2,\mu^2)}
{(k^2+\mu^2)} I'(k^2, \mu^2) + \fr{3}{2}\Delta^2(k^2,\mu^2)
I'(k^2, \mu^2)I''(k^2, \mu^2)
  \right] +
\\ 
&+&\left.\left[ \fr{1}{2} \Delta (k^2,\mu^2) I'''(k^2, \mu^2) -
 \fr{1}{2} 
\fr{1}{(4 \pi)^2 \mu^4}
\fr{\Delta (k^2,\mu^2)}{(k^2+\mu^2)} \right]\right\}\, .
\eeqn
All these terms have a one-to-one correspondance
  to the Feynman diagrams listed in
 ref.~\cite{dicecio}.

\begin{references}
\bibitem{landau}
L.D. Landau and I.Ya. Pomeranchuk, Dokl. Akad. Nauk. {\bf 102} (1955) 489;
L.D. Landau et al., Collected papers of L.D. Landau, (Gorden and Breach, 
New York, 1965).
\bibitem{wilsonkogut74}
K.G. Wilson and J. Kogut, Phys. Reports {\bf 12} (1974) 75.
\bibitem{root}
R.G. Root, Phys. Rev. {\bf D 10} (1974) 3322.
\bibitem{lautrup}
B. Lautrup, Phys. Lett. {\bf B 69} (1977) 109.
\bibitem{'tHooft}
G. 't Hooft, ``The Whys of Subnuclear Physics", Erice 1977, ed. by A.
Zichichi, Plenum Press, New York.
\bibitem{dingle}
R.B. Dingle, ``Asymptotic Expansions: their Derivation and Interpretation",
Academic Press, London, 1973;
G.N. Hardy, ``Divergent Series", Oxford Univ. Press 1949.
\bibitem{parisi}
G. Parisi, Phys. Lett. {\bf B 66} (1977) 382;
G. Parisi, Phys. Lett. {\bf B 76} (1978) 65;
G. Parisi, Phys. Reports {\bf 49} (1979) 215.
\bibitem{all}
A.H. Mueller, Nucl. Phys. {\bf B 250} (1985) 327;
M. Beneke, Phys. Lett. {\bf B 307} (1993) 154;
M. Beneke and V.I. Zakharov,  Phys. Lett. {\bf B 312} (1993) 340.
\bibitem{coleman}
S. Coleman, R. Jackiw and H. Politzer, Phys. Rev. {\bf D 10} (1974) 2491.
\bibitem{dicecio}
G. Di Cecio, Tesi di Dottorato, Universit\'a di Pisa, 1994;
G. Di Cecio and G. Paffuti, preprint IFUP-TH 48/93 (1993).
\end{references}
\end{document}